\documentclass[aps,prl,twocolumn,10pt,showpacs,superscriptaddress]{revtex4}

\usepackage{graphicx}
\usepackage{amsmath,units}
\usepackage{mathrsfs}
\usepackage{color}

\begin{document}

%============================================================
\title{Two Coupled Mechanisms Produce Fickian, yet non-Gaussian Diffusion in Heterogeneous Media  } 
%============================================================

\author{Indrani Chakraborty}

\affiliation{School of Chemistry, Tel Aviv
  University, Tel Aviv 6997801, Israel}

\author{Yael Roichman}
\email{roichman@tauex.tau.ac.il}

\affiliation{School of Chemistry, Tel Aviv
  University, Tel Aviv 6997801, Israel}
\affiliation{School of Physics \& Astronomy, Tel Aviv
	University, Tel Aviv 6997801, Israel}

\date{\today}

\begin{abstract}
Fickian yet non-Gaussian diffusion is observed in several biological and soft matter systems, yet the underlying mechanisms behind the emergence of non-Gaussianity while retaining a linear mean square displacement remain speculative. Here, we characterize quantitatively the effect of spatial heterogeneities on the appearance of non-Gaussianity in Fickian diffusion. We study the diffusion of fluorescent colloidal particles in a matrix of micropillars having a range of structural configurations: from completely ordered to completely random. We show that non-Gaussianity emerges as a direct consequence of two coupled factors; individual particle diffusivities become spatially dependent in a heterogeneous randomly structured environment, and the spatial distribution of the particles varies significantly in such environments, further influencing the diffusivity of a single particle. The coupled mechanisms lead to a considerable non-Gaussian nature even due to weak disorder in the arrangement of the micropillars. A simple mathematical model validates our hypothesis that non-Gaussian yet Fickian diffusion in our system arises from the superstatistical behavior of the ensemble in a structurally heterogeneous environment. The two mechanisms identified here are relevant for many systems of crowded heterogeneous environments where non-Gaussian diffusion is frequently observed, for example in biological systems, polymers, gels and porous materials.
\end{abstract}


\maketitle
%------------------------------------------------

Einstein’s theory of Brownian motion shows that for colloidal particles diffusing in two-dimensions in a simple fluid, the mean square displacement ({\em MSD}) is given by $ \text{\em MSD}  = 4D\tau$ where $D$ is the diffusion coefficient and $\tau$ is the lag time. The displacement distribution $G(\Delta x)$ for a Brownian particle is Gaussian. However, in many systems \cite{Orpe2007,Gollub1991,Bertrand2012,Chaudhuri2007,Stuhrmann2012,Wang2009,He2016,Grady2017,Witzel2019,Skaug2015,Wang2019,Chen2017,Dragulescu2002} such as  granular materials \cite{Orpe2007}, glassy materials \cite{Chaudhuri2007}, and biological systems \cite{Stuhrmann2012,Wang2009,He2016,Grady2017,Witzel2019}, a non-Gaussian $G(\Delta x$) has been observed. Usually, the non-Gaussian nature of $G(\Delta x$)  is related to a process of anomalous diffusion, with $ \text{{\em MSD}}  = 4\tilde D\tau^n$, $n$ being the diffusion exponent.  Surprisingly, in several cases diffusion has been observed to be Fickian but not Gaussian, that is the {\em MSD} remains linear in time, but $G(\Delta x$) is not Gaussian. This peculiar behavior has been referred to as `Anomalous, yet Brownian' or `Fickian yet non-Gaussian' diffusion (FNG), and has been observed in a wide variety of systems \cite{Leptos2009,Wang2009,Wang2012,Skaug2013,Yu2013,Chakraborty2019,Kurtuldu2011,Guan2014,Thorneywork2016}  ranging from tracer colloids diffusing in suspensions of swimming microorganisms \cite{Leptos2009} to polymer chains diffusing on a surface \cite{Skaug2013,Yu2013}. Several theories have been put forward to explain such behavior which include the ‘diffusing diffusivity’ model \cite{Chechkin2017,Chubynsky2014,Lanoiselee2018,Malgaretti2016} that considers dynamic heterogeneities experienced by each colloidal particle in a changing environment. A second suggestion \cite{Wang2009,Wang2012,Guan2014,Chechkin2017} is that such motion arises from the ‘superstatistical’ behavior of an ensemble with each member having different diffusive parameters in a spatially varying environment. Guan \textit{et al.} \cite{Guan2014} observed FNG diffusion in a system of probe colloidal particles diffusing in a static matrix of densely packed bigger particles. It was postulated that the differences in the local configurations of the larger matrix particles led to the observed Fickian yet non Gaussian diffusion. However, since the configuration of the matrix particles could not be controlled experimentally, it was not possible to quantitatively estimate the effect of the spatial heterogeneities on the non-Gaussian nature of the diffusion.\par 
Here we study the emergence of non-Gaussianity from the structural randomness of the environment and quantify the percentage of disorder that gives rise to such non-Gaussian yet Fickian diffusion. This is achieved by fabricating arrays of micropillars with different degrees of randomness and tracking the motion of colloidal particles through them. We find that even a small degree of randomness leads to extensive non-Gaussianity due to two coupled mechanisms arising directly from the structural randomness in the micropillar arrays. Our experiments establish the importance of structural disorder in FNG diffusion.\par
Our samples consist of micropillars (cylindrical cross-section with diameter = 6 $\mu m$, height = 6 $\mu m$) made of the photoresist SU8 using standard photolithography techniques on glass cover slips. Samples with different area fractions and degrees of randomness were fabricated by designing corresponding masks (see Supplemental Material \cite{Supple}). A flat glass control was made by exposing and developing an SU8 coated cover-slip without introducing any mask. A suspension of fluorescent polystyrene particles ($4.19 \pm 0.27$~$\mu m$ in diameter, Bangs Laboratories Inc.) was placed onto the micropillar arrays and sealed with a top glass plate. Particles were allowed to sediment to the bottom of the chamber and diffuse there. The area fraction of the polystyrene particles on the glass surface was $\approx$ 6\%.  The diffusion of the probe particles was imaged (Olympus IX-71 microscope, 40X objective) in both bright field and fluorescence modes and tracked at 50 fps using conventional video microscopy \cite{Allan2018,Crocker1996}. \par
\begin{figure}[h!]
	\centering
	\includegraphics[scale=0.63]{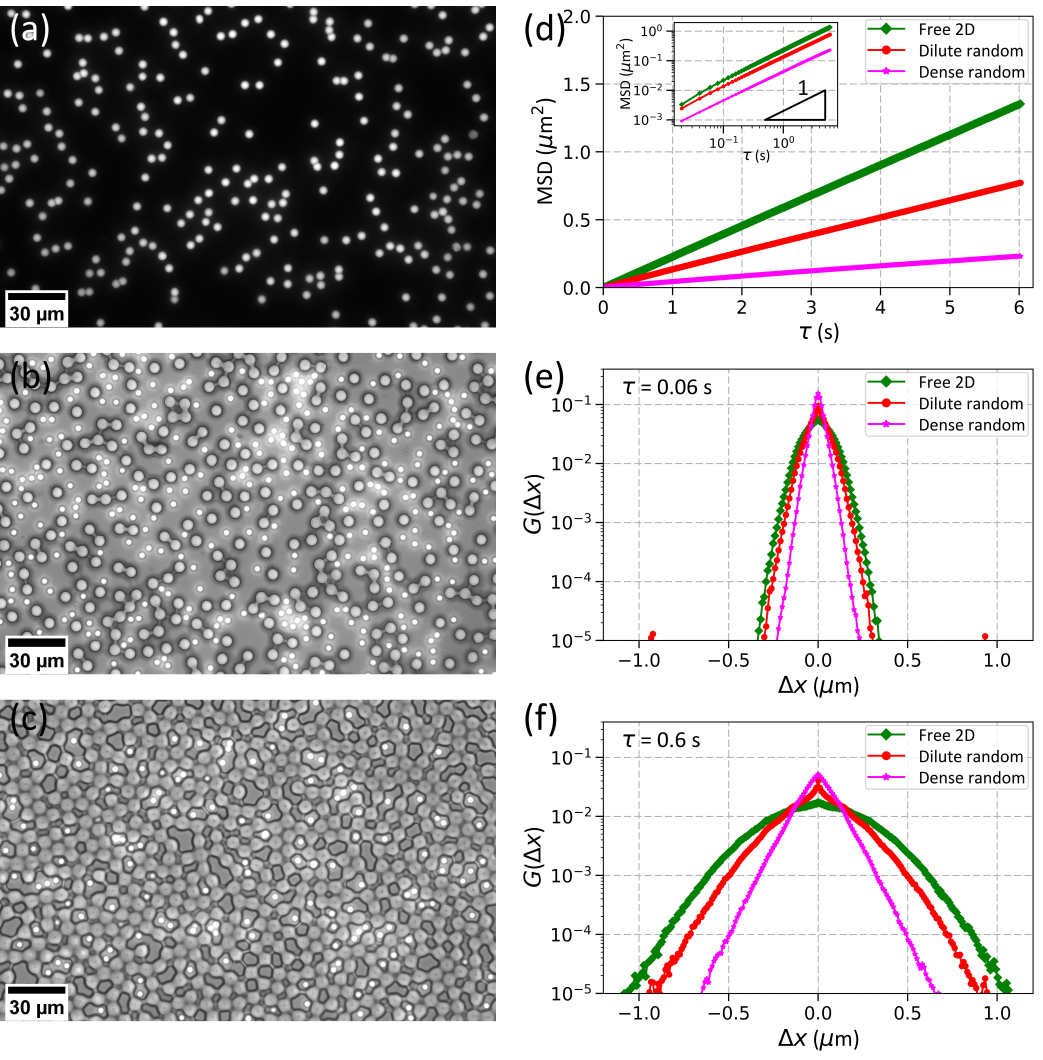}
	\caption{\label{fig:Fig1} Increase in density of the micropillars makes diffusion Fickian yet non-Gaussian. Microscope images  in transmission mode of fluorescent tracer particles (bright spheres) diffusing on (a) a glass surface (b) in a matrix containing an area fraction of $\phi=20\%$  of randomly placed micropillars (bigger circles), i.e., `dilute random' (c) in a matrix containing randomly placed micropillars at $\phi=72\%$, i.e., `dense random'.  The time ensemble averaged {\em MSD} plots are shown in (d) while inset shows the log-log plots for the same systems indicating clear Fickian diffusion with $n = 1$. $G(\Delta x)$ plots are shown for lag times (e) $\tau = 0.06$~s and (f) $\tau = 0.6$~s. Note the distinct exponential behavior of the system with the dense random micropillars while the {\em MSD} is still clearly linear in time. }
\end{figure}

Initially we consider two cases of randomly arranged arrays of micropillars along with a control sample of polystyrene spheres diffusing in 2D on a planar glass surface (Fig. \ref{fig:Fig1}(a)). In the first sample the micropillars have an area fraction of $\phi\approx20\%$ referred to as `dilute random' (Fig. \ref{fig:Fig1}(b)) while in the second case, the area fraction is $\phi\approx$ 72\% referred to as `dense random' (Fig. \ref{fig:Fig1}(c)). It should be noted that in the dense random case, the randomly placed micropillars were so close to each other that many of them were connected. This led to the formation of pockets separated by rigid walls in which particles could get trapped. The {\em MSD} of the particles in all three cases was observed to be linear, i.e., $n = 1$, over a time scale spanning three decades (Fig. \ref{fig:Fig1}(d)).  In contrast, $G(\Delta x$) is Gaussian only for the free diffusion samples and a markedly non-Gaussian distribution with a small peak centered at zero was obtained for all other samples  (Fig. \ref{fig:Fig1}(e) and (f)).  For the dense random case, the change in $G(\Delta x$) was even more dramatic having a clear exponential appearance. It should be noted here that at sufficiently long time intervals (lag time $> 5$ s), the dense random sample showed $n = 0.8$ instead of 1, indicating a caging effect at these time scales. However, in the time scale in which the diffusion is Fickian, $G(\Delta x$) clearly exhibits an exponential behavior. This indicates that an increase in area fraction of the micropillars leads to greater non-Gaussianity, in agreement with a previous report \cite{He2014} which showed that increasing confinement leads to a larger non-Gaussianity in diffusion.

\begin{figure}[h!]
	\centering
	\includegraphics[scale=0.63]{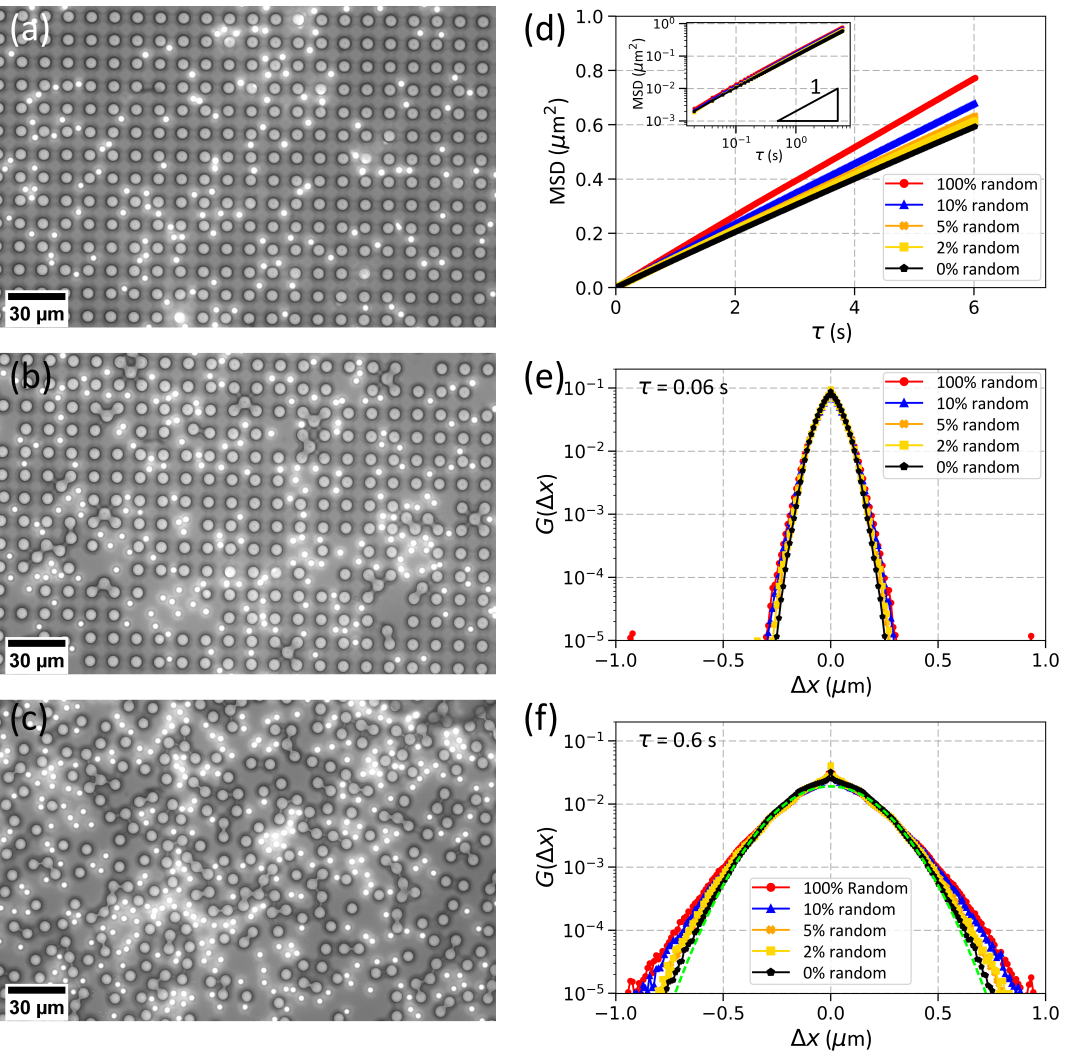}
	\caption{\label{fig:Fig2} Increase in randomness of the micropillars makes diffusion Fickian yet non-Gaussian. Fluorescent silica particles diffusing in a matrix of micropillars at $\phi=20\%$ in (a) completely ordered (b) semi-random (10\% randomness) and (c) completely random (100\% randomness) arrangements. The {\em MSD} plots for randomness values of 0\%, 2\%, 5\%, 10\% and 100\% are shown in (d) while inset shows the log-log plot. $G(\Delta x)$ distribution is shown for lag times (e) $\tau = 0.06 s$ and (f) $\tau = 0.6 s$. The dashed green line in (f) shows a Gaussian fit to the $G(\Delta x)$ plot for the 100\% random sample. Note the increasing non-Gaussian behavior with the increase of randomness even at same area fraction.}
\end{figure}

\begin{figure}[h!]
	\centering
	\hspace*{0cm}\includegraphics[scale=0.53]{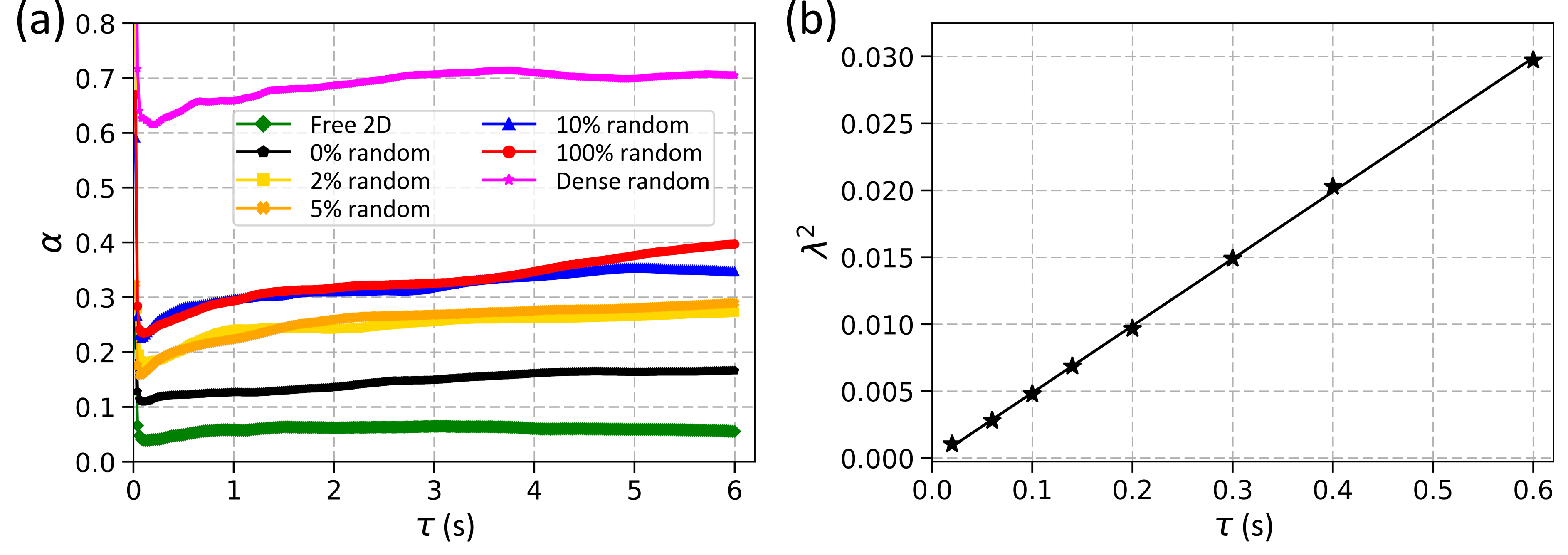}
	\caption{\label{fig:Fig3}  Measurement of non-Gaussianity as a function of the randomness of the micropillars. (a) The non-Gaussian parameter for the seven different cases: free 2D, 0\%, 2\%, 5\%, 10\%, 100\% random and dense random. (b) The value of the factor $\lambda^2$  obtained as a function of lag time is shown in (b). Note the excellent linear fit. }
\end{figure}

To explore the effect of the structural heterogeneity of the micropillars on the diffusion of the probe particles, we made a series of five samples each having $\phi= 20\%$  of micropillars (i.e., `dilute random') with micropillars arrangements ranging from completely ordered to completely random. The samples were:  a) `ordered' with 0\% randomness (square array with center to center distance of 12 $\mu m$) (Fig. \ref{fig:Fig2}(a)), `semi-random' including samples with b) 2\%, c) 5\% and d) 10\% randomness (Fig. \ref{fig:Fig2}(b)), and (e) a 100\% random sample (Fig. \ref{fig:Fig2}(c)). The semi-random samples were obtained by choosing a given percentage of the micropillars randomly and then shifting them by a random amount from their positions in the perfectly ordered structure while designing the mask. This is essentially equivalent to the introduction of a given percentage of `defects' in an ordered structure. All samples exhibited Fickian diffusion (Fig. \ref{fig:Fig2}(d)) with an increase in the diffusion constant as the structure randomness increased. However, $G(\Delta x$) became less Gaussian as structural randomness increased (Fig.~\ref{fig:Fig2}(e) and (f)).We note that even with an inclusion of only 10\% randomness in the structure, the $G(\Delta x$) plots are very similar to the case for the 100\% random structure, indicating very high sensitivity of the system to structural`defects'. This behavior is also reflected in the non-Gaussian parameter $\alpha$  \cite{Guan2014}, where  $\alpha(t)=\frac{\langle\Delta {x^4}(t)\rangle}{3{{\langle\Delta {x^2}(t)\rangle}^2}}-1$ (Fig. \ref{fig:Fig3}(a)). $\alpha$ is very close to zero for the free 2D diffusion case, while it is $\approx$ 1.2 times higher than that of the ordered micropillars for the 2\% and 5\% samples, and is nearly the same as for the 10\% and 100\% disordered samples. Notably, $\alpha$ remains nearly constant over several lag times for most cases. For the dense random case $\alpha$ is $\approx$ 5 times that of the ordered case, indicating a very high degree of non-Gaussianity. We also verified the exponential nature of $G(\Delta x$) for the dense random case \cite{Wang2009} by fitting the equation $G(\Delta x)=A{e^{-\frac{|x|}{\lambda}}}$ to $G(\Delta x$) plots over several different lag times and calculated the corresponding values of $\lambda$. From Fig. \ref{fig:Fig3}(b) we see that indeed ${\lambda^2}\sim\tau$, confirming the Fickian, yet exponential behavior. Note that emergence of non-Gaussianity with randomness is also observed at higher area fractions (28.3\%) of micropillars (see Supplemental Material Fig. S1 \cite{Supple}).

\begin{figure}[h!]
	\centering
	\hspace*{-1cm}\includegraphics[scale=0.58]{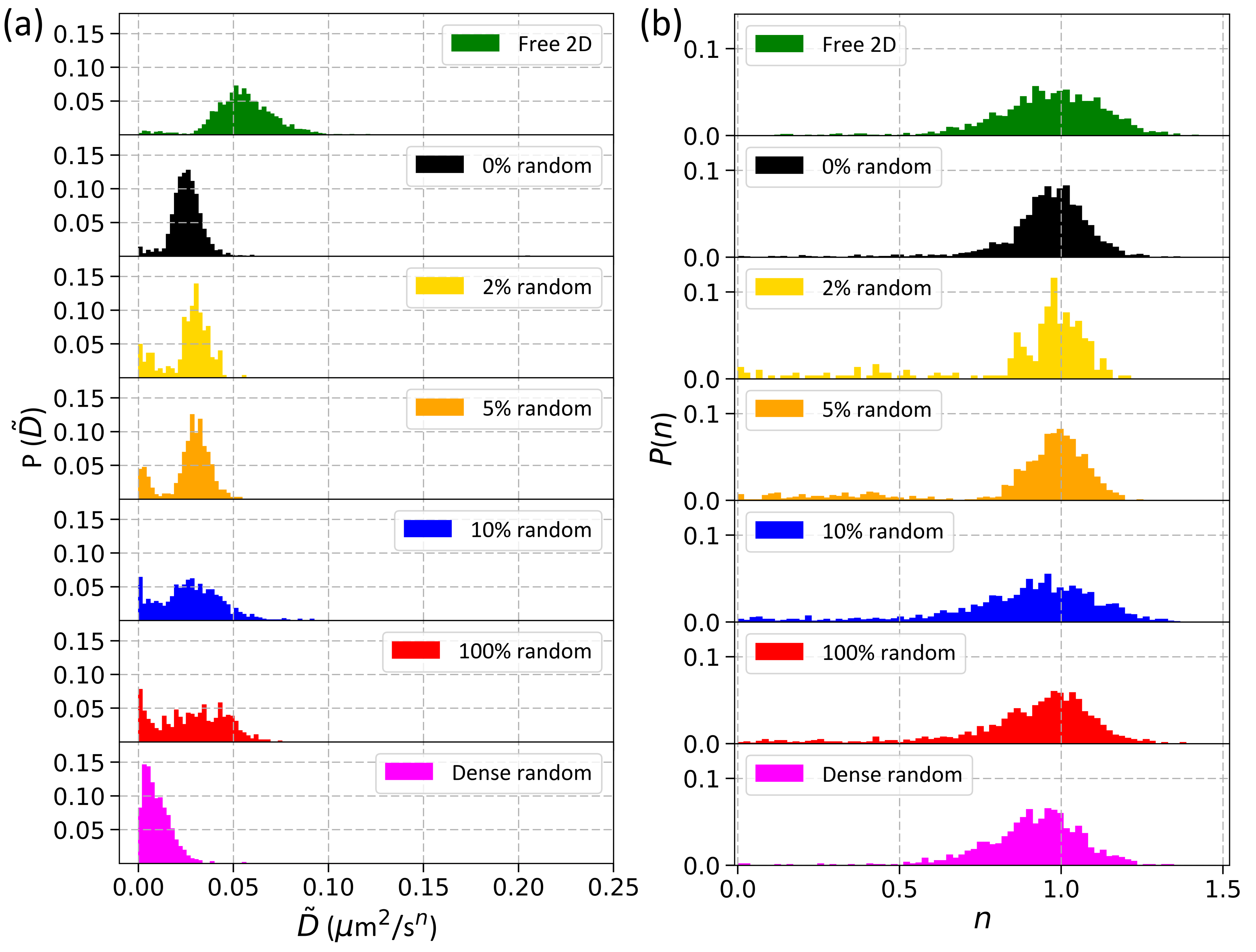}
	\caption{\label{fig:Fig4} (a) Probability distribution $P(\tilde D)$ of the diffusion constant $(\tilde D)$ for the seven cases: free 2D, 0\%, 2\%, 5\%, 10\% and 100\% randomness and dense random respectively. Note the existence of two populations in the $P(\tilde D)$ plots for the dilute random samples with different degrees of randomness while the dense random sample shows a near-exponential behavior. (b) Probability distribution $P(n)$ of the diffusion exponents $(n)$ calculated from the linear fits of log-log {\em MSD} plots for each individual particle. Note that the distribution peaks about $n = 1.0$, indicating predominantly normal diffusion for the majority of the particles for all the cases.}
\end{figure}

\begin{figure}[h!]
	\centering
	\includegraphics[scale=0.63]{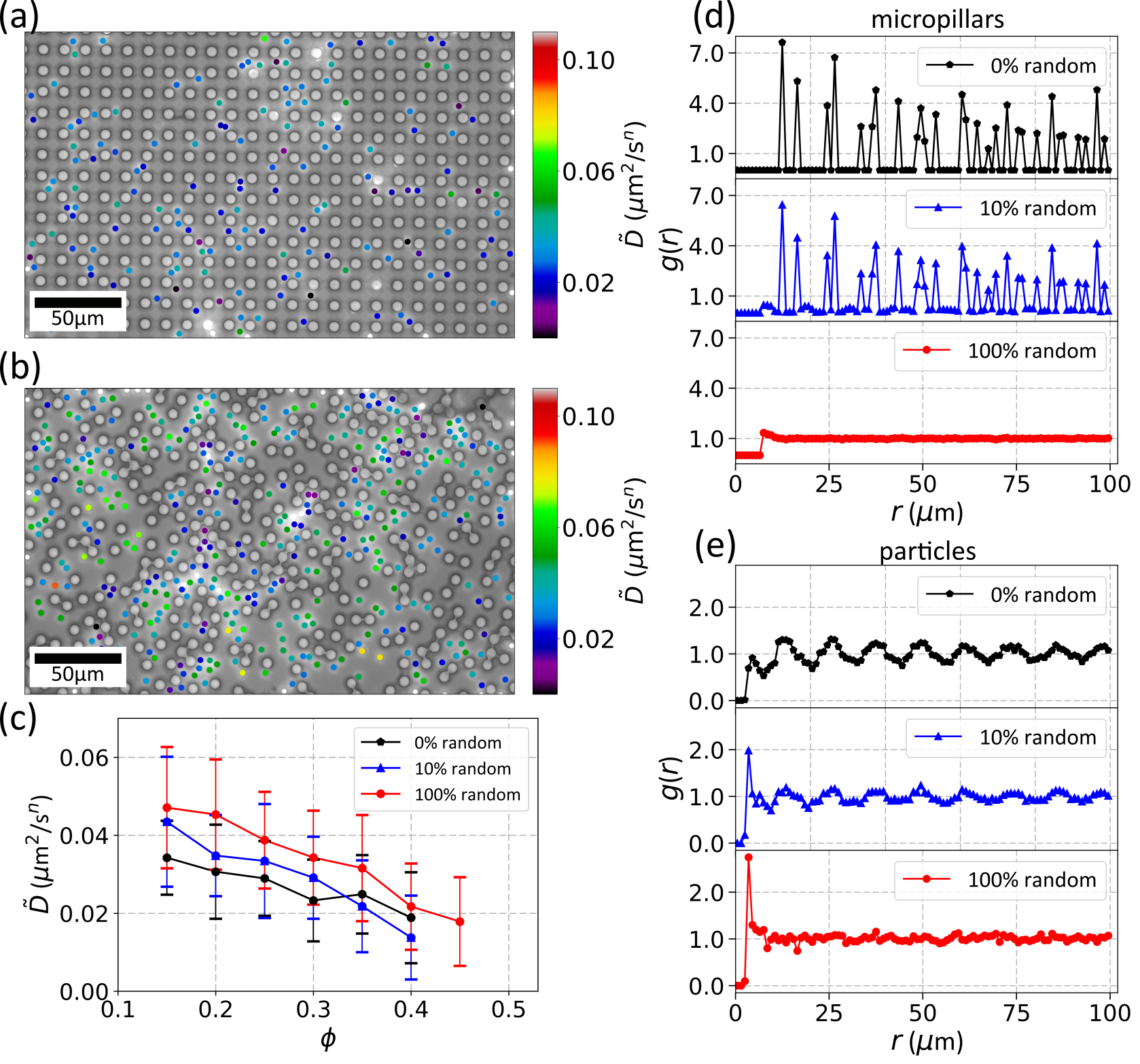}
	\caption{\label{fig:Fig5} 2D color maps representing the $\tilde D$ values of colloids for (a) 0\% random and (b) 100\% random micropillars. Note the higher $\tilde D$ values for particles in open spaces and lower $\tilde D$ values for particles close to a wall or other particles for the random sample. A plot showing the decrease of average $\tilde D$ with increasing local area fraction $\phi_{loc}$ (composed of micropillars and neighbouring particles) around each given particle within a radius of interest of 50 pixels is shown in (c). Pair correlation function $g(r)$ of the (d) micropillars and (e) particles for the 0\% random, 10\% random and 100\% random cases.}
\end{figure}

To understand the mechanism behind the increased non-Gaussianity due to density and structural randomness, we focus on the probability distribution of the diffusion constants $P(\tilde D)$ over the ensemble for the seven systems: free 2D, `dilute random' samples with randomness of 0\%, 2\%, 5\%, 10\% and 100\%, and the dense random case (Fig. \ref{fig:Fig4}(a)). The diffusion constants here are calculated from the intercepts of the individual particle MSDs in log-log plots (see Supplemental Material \cite{Supple}). For all the samples with different degrees of randomness (2\%-100\%) $P(\tilde D)$ is not Gaussian, and has two notable features: a) a second peak at lower values of the diffusion constant $(\tilde D)$ appears, indicating that a portion of the ensemble moves very slowly b) there is an extended tail at higher values of $\tilde D$ indicating that a portion of the ensemble moves faster than the average. For the dense random sample, the distribution looks almost exponential. In spite of such wide differences in the $P(\tilde D)$ distributions, the single particle diffusion exponent distribution, $P(n)$, peaks at $n = 1.0$ for all the seven cases. This indicates Fickian diffusion on the single particle level for the great majority of the particles (Fig. \ref{fig:Fig4}(b)). Further analysis shows that with an increasing degree of randomness, the small fraction of particles with $n$ $\neq$ 1 increases, along with an increase in the fraction of both slow and fast moving particles. (see Supplemental Material Fig. S2(a)-(b) \cite{Supple}).\par
 
The explicit difference in $P(\tilde D)$ for the random and ordered micropillars, therefore, points to the fact that the local spatial heterogeneities in the random structure result in the ensemble having subgroups of particles that see different local environments and consequently diffuse differently. In a random structure there are more open areas and narrow cavities as compared to an ordered structure. A particle can get trapped in the cavities surrounded by the micropillar walls on one hand, while on the other hand it might happen to be in a location which is far off from any micropillar wall. While the former leads to a slower motion or even caged diffusion, in the later, the particle can diffuse unhindered. This can be seen in Fig. \ref{fig:Fig5}(a)-(b) where we show the  $\tilde D$  values per particle for the 0\% and 100\% random samples in relation to their position in the matrix.  An increase in the local area fraction $\phi_{loc}$ of neighbouring features (both micropillars and particles) around each particle therefore leads to a lowering of the average  $(\tilde D)$ values as seen in Fig. \ref{fig:Fig5}(c). Another indication that the non-Gaussian distribution results from local density variations rather than the long range order of the ordered sample is that the hydrodynamic interactions between distant tracer particles is similar for ordered and disordered samples at intermediate and long range (see Fig. S3 of Supplemental Material \cite{Supple}).  The spatial distribution of the probe particles themselves is affected by structural randomness varying more significantly in random samples. As a result, the diffusion of a particle entrapped in the crowded regions gets slowed down while diffusion is faster in regions with very few particles. Comparing the 2D pair correlation function $g(r)$ for both the micropillars (Fig. \ref{fig:Fig5}(d)) and the particles (Fig. \ref{fig:Fig5}(e)) we see that for the ordered sample, $g(r)$ for the particles resembles that of the micropillars in the sense that both show periodic behaviors with peaks at 12 $\mu m$, 24 $\mu m$, 36 $\mu m$… etc. In contrast, $g(r)$ for the 100\% random case saturates very fast to 1.0, indicating a wide distribution of the nearest neighbor distances. Also the peak at 6 $\mu m$ is the highest for the particles in the random structure as compared to the very small peak in the ordered structure (Fig. \ref{fig:Fig5}(e)), indicating the presence of particles at very close distances to other particles or very crowded regions for the random arrangement. The amplified effect of the two coupled factors - the location dependence of the diffusivity values in a spatially heterogeneous structure, and slowing or speeding up of diffusion near crowded or free regions, produces the extensive non-Gaussian response even at a dilute concentration of defects. The non-Gaussianity persists at both short and long time scales as seen from the nearly constant behavior of the non-Gaussian parameter over more than two decades of lag time (where the smallest time interval is 0.02 s). This is because each individual particle does not move appreciably from its location within our experimental time frame, and does not experience the whole set of environments in the sample. In contrast, in the `diffusing diffusivity' model, each individual particle experiences a slowly fluctuating environment over their trajectory and the non-Gaussian behavior is seen for each particle. In that model, at longer time scales the diffusion of a single particle transitions to Gaussian behavior \cite{Wang2009,Chechkin2017}.\par

To verify our hypothesis, we did a simple calculation. Assuming a Gaussian $G(\Delta x$) for each particle, we took the weights of the different diffusion constant values from our measurements (Fig. 4(a)) and summed over the individual displacements using $G(\Delta x,t)= \Sigma_{i=1}^N \frac{w_i}{\sqrt{4\pi \tilde D_i t}} e^{-\frac{x^2}{4\tilde D_i t}}$  where $w_i$ is the weight of each given value of the diffusion constant $\tilde D_i$. Our calculation results at a lag time of 0.6 s (see Supplemental Material Fig. S4(a)-(g)) \cite{Supple} agree well with our $G(\Delta x$) measurements, indicating that indeed the spread in the diffusivities arising from the local spatial heterogeneities in a random structure is the reason behind the FNG diffusion in these systems. This is in essence superstatistical behavior where the fast, random motion of the colloidal particles is superposed with the variations in the environment with specific areas having higher and lower diffusivity values \cite{Chechkin2017}. Previously, a diffusion constant distribution of the form $P(D)=  \frac{1}{\langle D \rangle} e^{-\frac{D}{\langle D \rangle}}$ where $\langle D \rangle$ is the mean diffusion constant was shown to produce an exponential $G(\Delta x$) \cite{Chubynsky2014, Hapca2009}. In our system, this is analogous to the dense random case.\par 

In conclusion we experimentally showed that the existence of spatial heterogeneities is a profoundly important factor behind the emergence of non-Gaussianity in Fickian diffusion. Both the area fraction and structural randomness of the heterogeneities contribute directly to increasing the non-Gaussianity culminating in exponential displacement distributions in extreme cases. Even weak disorder in the system (10\%) can produce an extensive deviation from Gaussianity. This extreme sensitivity to randomness in the structure originates from the amplified contributions of two coupled effects; increased density variations in the matrix structure and in the tracer particle distribution. This combination leads to a wide range of diffusivities and produces the superstatistical non-Gaussian behavior for Fickian diffusion. The importance of our study lies in the fact that it is the first controlled experiment quantitatively examining the effect of environmental heterogeneities on the nature of diffusion of colloidal particles. We expect this work to be an important addition in the endeavor to understand non-conventional diffusive behaviors in complex systems. Our findings are universally relevant in systems in which non-Gaussianity arises explicitly from the existence of spatio-temporal heterogeneities, for example, in diffusion of nutrients across the cell cytoplasm containing a sea of biomacromolecules \cite{Witzel2019}, diffusion in protein crowded lipid bilayers \cite{Jeon2016}, diffusion of dust particles in 2D equilibrium Yukawa liquids \cite{Ghannad2019}, liposomes diffusing in nematic solutions of F-actin filaments \cite{Wang2012}, colloidal particles diffusing among swimming cells \cite{Kurtuldu2011} or in a dilute field of optical traps \cite{Mora2018} and swimming microorganisms navigating through colloidal particles \cite{Jeanneret2016}.

This research was supported by the Israel Science Foundation (grant No. 988/17) and the PBC Post-doctoral fellowship of the Council for Higher education, Israel.

%\bibliography{Obstacles}

\begin{thebibliography}{26}
	\expandafter\ifx\csname natexlab\endcsname\relax\def\natexlab#1{#1}\fi
	\expandafter\ifx\csname bibnamefont\endcsname\relax
	\def\bibnamefont#1{#1}\fi
	\expandafter\ifx\csname bibfnamefont\endcsname\relax
	\def\bibfnamefont#1{#1}\fi
	\expandafter\ifx\csname citenamefont\endcsname\relax
	\def\citenamefont#1{#1}\fi
	\expandafter\ifx\csname url\endcsname\relax
	\def\url#1{\texttt{#1}}\fi
	\expandafter\ifx\csname urlprefix\endcsname\relax\def\urlprefix{URL }\fi
	\providecommand{\bibinfo}[2]{#2}
	\providecommand{\eprint}[2][]{\url{#2}}
	
	\bibitem[{\citenamefont{Orpe and Kudrolli}(2007)}]{Orpe2007}
	\bibinfo{author}{\bibfnamefont{A.~V.} \bibnamefont{Orpe}} \bibnamefont{and}
	\bibinfo{author}{\bibfnamefont{A.}~\bibnamefont{Kudrolli}},
	\bibinfo{journal}{Phys. Rev. Lett.} \textbf{\bibinfo{volume}{98}},
	\bibinfo{pages}{238001} (\bibinfo{year}{2007}).
	
	\bibitem[{\citenamefont{Gollub et~al.}(1991)\citenamefont{Gollub, Clarke,
			Gharib, Lane, and Mesquita}}]{Gollub1991}
	\bibinfo{author}{\bibfnamefont{J.~P.} \bibnamefont{Gollub}},
	\bibinfo{author}{\bibfnamefont{J.}~\bibnamefont{Clarke}},
	\bibinfo{author}{\bibfnamefont{M.}~\bibnamefont{Gharib}},
	\bibinfo{author}{\bibfnamefont{B.}~\bibnamefont{Lane}}, \bibnamefont{and}
	\bibinfo{author}{\bibfnamefont{O.~N.} \bibnamefont{Mesquita}},
	\bibinfo{journal}{Phys. Rev. Lett.} \textbf{\bibinfo{volume}{67}},
	\bibinfo{pages}{3507} (\bibinfo{year}{1991}).
	
	\bibitem[{\citenamefont{Bertrand et~al.}(2012)\citenamefont{Bertrand, Fygenson,
			and Saleh}}]{Bertrand2012}
	\bibinfo{author}{\bibfnamefont{O.~J.~N.} \bibnamefont{Bertrand}},
	\bibinfo{author}{\bibfnamefont{D.~K.} \bibnamefont{Fygenson}},
	\bibnamefont{and} \bibinfo{author}{\bibfnamefont{O.~A.} \bibnamefont{Saleh}},
	\bibinfo{journal}{Proc. Natl. Acad. Sci.} \textbf{\bibinfo{volume}{109}},
	\bibinfo{pages}{17342} (\bibinfo{year}{2012}).
	
	\bibitem[{\citenamefont{Chaudhuri et~al.}(2007)\citenamefont{Chaudhuri,
			Berthier, and Kob}}]{Chaudhuri2007}
	\bibinfo{author}{\bibfnamefont{P.}~\bibnamefont{Chaudhuri}},
	\bibinfo{author}{\bibfnamefont{L.}~\bibnamefont{Berthier}}, \bibnamefont{and}
	\bibinfo{author}{\bibfnamefont{W.}~\bibnamefont{Kob}},
	\bibinfo{journal}{Phys. Rev. Lett.} \textbf{\bibinfo{volume}{99}},
	\bibinfo{pages}{060604} (\bibinfo{year}{2007}).
	
	\bibitem[{\citenamefont{Skaug et~al.}(2015)\citenamefont{Skaug, Wang, Ding, and
			Schwartz}}]{Skaug2015}
	\bibinfo{author}{\bibfnamefont{M.~J.} \bibnamefont{Skaug}},
	\bibinfo{author}{\bibfnamefont{L.}~\bibnamefont{Wang}},
	\bibinfo{author}{\bibfnamefont{Y.}~\bibnamefont{Ding}}, \bibnamefont{and}
	\bibinfo{author}{\bibfnamefont{D.~K.} \bibnamefont{Schwartz}},
	\bibinfo{journal}{ACS Nano} \textbf{\bibinfo{volume}{9}},
	\bibinfo{pages}{2148} (\bibinfo{year}{2015}).
	
	\bibitem[{\citenamefont{Wang et~al.}(2019)\citenamefont{Wang, Wu, Liu, Chen,
			and Schwartz}}]{Wang2019}
	\bibinfo{author}{\bibfnamefont{D.}~\bibnamefont{Wang}},
	\bibinfo{author}{\bibfnamefont{H.}~\bibnamefont{Wu}},
	\bibinfo{author}{\bibfnamefont{L.}~\bibnamefont{Liu}},
	\bibinfo{author}{\bibfnamefont{J.}~\bibnamefont{Chen}}, \bibnamefont{and}
	\bibinfo{author}{\bibfnamefont{D.~K.} \bibnamefont{Schwartz}},
	\bibinfo{journal}{Phys. Rev. Lett.} \textbf{\bibinfo{volume}{123}},
	\bibinfo{pages}{118002} (\bibinfo{year}{2019}).
	
	\bibitem[{\citenamefont{Chen et~al.}(2017)\citenamefont{Chen,Qian and Lu}}]{Chen2017}
	\bibinfo{author}{\bibfnamefont{T.} \bibnamefont{Chen}},
	\bibinfo{author}{\bibfnamefont{H.~J.} \bibnamefont{Qian}} \bibnamefont{and}
	\bibinfo{author}{\bibfnamefont{Z.~Y.}~\bibnamefont{Lu}}, 
	\bibinfo{journal}{Chem. Phys. Lett.} \textbf{\bibinfo{volume}{687}},
	\bibinfo{pages}{96} (\bibinfo{year}{2017}).
	
	\bibitem[{\citenamefont{Drăgulescu and Yakovenko}(2002)}]{Dragulescu2002}
	\bibinfo{author}{\bibfnamefont{A.~A.} \bibnamefont{Drăgulescu}}
	\bibnamefont{and} \bibinfo{author}{\bibfnamefont{V.~M.}
		\bibnamefont{Yakovenko}}, \bibinfo{journal}{Quant. Financ.}
	\textbf{\bibinfo{volume}{2}}, \bibinfo{pages}{443} (\bibinfo{year}{2002}).
	
	\bibitem[{\citenamefont{Stuhrmann et~al.}(2012)\citenamefont{Stuhrmann, {Soares
				e Silva}, Depken, MacKintosh, and Koenderink}}]{Stuhrmann2012}
	\bibinfo{author}{\bibfnamefont{B.}~\bibnamefont{Stuhrmann}},
	\bibinfo{author}{\bibfnamefont{M.}~\bibnamefont{{Soares e Silva}}},
	\bibinfo{author}{\bibfnamefont{M.}~\bibnamefont{Depken}},
	\bibinfo{author}{\bibfnamefont{F.~C.} \bibnamefont{MacKintosh}},
	\bibnamefont{and} \bibinfo{author}{\bibfnamefont{G.~H.}
		\bibnamefont{Koenderink}}, \bibinfo{journal}{Phys. Rev. E}
	\textbf{\bibinfo{volume}{86}}, \bibinfo{pages}{020901(R)}
	(\bibinfo{year}{2012}).
	
	\bibitem[{\citenamefont{Wang et~al.}(2009)\citenamefont{Wang, Anthony, Bae, and
			Granick}}]{Wang2009}
	\bibinfo{author}{\bibfnamefont{B.}~\bibnamefont{Wang}},
	\bibinfo{author}{\bibfnamefont{S.~M.} \bibnamefont{Anthony}},
	\bibinfo{author}{\bibfnamefont{S.~C.} \bibnamefont{Bae}}, \bibnamefont{and}
	\bibinfo{author}{\bibfnamefont{S.}~\bibnamefont{Granick}},
	\bibinfo{journal}{Proc. Natl. Acad. Sci.}
	\textbf{\bibinfo{volume}{106}}, \bibinfo{pages}{15160}
	(\bibinfo{year}{2009}).
	
	\bibitem[{\citenamefont{He et~al.}(2016)\citenamefont{He, Song, Su, Geng,
			Ackerson, Peng, and Tong}}]{He2016}
	\bibinfo{author}{\bibfnamefont{W.}~\bibnamefont{He}},
	\bibinfo{author}{\bibfnamefont{H.}~\bibnamefont{Song}},
	\bibinfo{author}{\bibfnamefont{Y.}~\bibnamefont{Su}},
	\bibinfo{author}{\bibfnamefont{L.}~\bibnamefont{Geng}},
	\bibinfo{author}{\bibfnamefont{B.~J.} \bibnamefont{Ackerson}},
	\bibinfo{author}{\bibfnamefont{H.~B.} \bibnamefont{Peng}}, \bibnamefont{and}
	\bibinfo{author}{\bibfnamefont{P.}~\bibnamefont{Tong}},
	\bibinfo{journal}{Nat. Commun.} \textbf{\bibinfo{volume}{7}},
	\bibinfo{pages}{11701} (\bibinfo{year}{2016}).
	
	\bibitem[{\citenamefont{Grady et~al.}(2017)\citenamefont{Grady, Parrish,
			Caporizzo, Seeger, Composto, and Eckmann}}]{Grady2017}
	\bibinfo{author}{\bibfnamefont{M.~E.} \bibnamefont{Grady}},
	\bibinfo{author}{\bibfnamefont{E.}~\bibnamefont{Parrish}},
	\bibinfo{author}{\bibfnamefont{M.~A.} \bibnamefont{Caporizzo}},
	\bibinfo{author}{\bibfnamefont{S.~C.} \bibnamefont{Seeger}},
	\bibinfo{author}{\bibfnamefont{R.~J.} \bibnamefont{Composto}},
	\bibnamefont{and} \bibinfo{author}{\bibfnamefont{D.~M.}
		\bibnamefont{Eckmann}}, \bibinfo{journal}{Soft Matter}
	\textbf{\bibinfo{volume}{13}}, \bibinfo{pages}{1873} (\bibinfo{year}{2017}).
	
	\bibitem[{\citenamefont{Witzel et~al.}(2019)\citenamefont{Witzel, G{\"{o}}tz,
			Lanoisel{\'{e}}e, Franosch, Grebenkov, and Heinrich}}]{Witzel2019}
	\bibinfo{author}{\bibfnamefont{P.}~\bibnamefont{Witzel}},
	\bibinfo{author}{\bibfnamefont{M.}~\bibnamefont{G{\"{o}}tz}},
	\bibinfo{author}{\bibfnamefont{Y.}~\bibnamefont{Lanoisel{\'{e}}e}},
	\bibinfo{author}{\bibfnamefont{T.}~\bibnamefont{Franosch}},
	\bibinfo{author}{\bibfnamefont{D.~S.} \bibnamefont{Grebenkov}},
	\bibnamefont{and} \bibinfo{author}{\bibfnamefont{D.}~\bibnamefont{Heinrich}},
	\bibinfo{journal}{Biophys. J.} \textbf{\bibinfo{volume}{117}},
	\bibinfo{pages}{203} (\bibinfo{year}{2019}).
	
	\bibitem{Leptos2009}
	K. C. Leptos, J. S. Guasto, J. P. Gollub, A. I. Pesci and R. E. Goldstein, Phys. Rev. Lett. {\bf 103}, 198103 (2009). 
	
	\bibitem[{\citenamefont{Wang et~al.}(2012)\citenamefont{Wang, Kuo, Bae, and
			Granick}}]{Wang2012}
	\bibinfo{author}{\bibfnamefont{B.}~\bibnamefont{Wang}},
	\bibinfo{author}{\bibfnamefont{J.}~\bibnamefont{Kuo}},
	\bibinfo{author}{\bibfnamefont{S.~C.} \bibnamefont{Bae}}, \bibnamefont{and}
	\bibinfo{author}{\bibfnamefont{S.}~\bibnamefont{Granick}},
	\bibinfo{journal}{Nat. Mater.} \textbf{\bibinfo{volume}{11}},
	\bibinfo{pages}{481}
	(\bibinfo{year}{2012}).
	
	\bibitem[{\citenamefont{Chakraborty et~al.}(2019)\citenamefont{Chakraborty,
			Rahamim, Avinery, Roichman, and Beck}}]{Chakraborty2019}
	\bibinfo{author}{\bibfnamefont{I.}~\bibnamefont{Chakraborty}},
	\bibinfo{author}{\bibfnamefont{G.}~\bibnamefont{Rahamim}},
	\bibinfo{author}{\bibfnamefont{R.}~\bibnamefont{Avinery}},
	\bibinfo{author}{\bibfnamefont{Y.}~\bibnamefont{Roichman}}, \bibnamefont{and}
	\bibinfo{author}{\bibfnamefont{R.}~\bibnamefont{Beck}},
	\bibinfo{journal}{Nano Lett.} \textbf{\bibinfo{volume}{19}},
	\bibinfo{pages}{6524} (\bibinfo{year}{2019}).
	
	\bibitem[{\citenamefont{Skaug et~al.}(2013)\citenamefont{Skaug, Mabry, and
			Schwartz}}]{Skaug2013}
	\bibinfo{author}{\bibfnamefont{M.~J.} \bibnamefont{Skaug}},
	\bibinfo{author}{\bibfnamefont{J.}~\bibnamefont{Mabry}}, \bibnamefont{and}
	\bibinfo{author}{\bibfnamefont{D.~K.} \bibnamefont{Schwartz}},
	\bibinfo{journal}{Phys. Rev. Lett.} \textbf{\bibinfo{volume}{110}},
	\bibinfo{pages}{256101} (\bibinfo{year}{2013}).
	
	\bibitem[{\citenamefont{Yu et~al.}(2013)\citenamefont{Yu, Guan, Chen, Bae, and
			Granick}}]{Yu2013}
	\bibinfo{author}{\bibfnamefont{C.}~\bibnamefont{Yu}},
	\bibinfo{author}{\bibfnamefont{J.}~\bibnamefont{Guan}},
	\bibinfo{author}{\bibfnamefont{K.}~\bibnamefont{Chen}},
	\bibinfo{author}{\bibfnamefont{S.~C.} \bibnamefont{Bae}}, \bibnamefont{and}
	\bibinfo{author}{\bibfnamefont{S.}~\bibnamefont{Granick}},
	\bibinfo{journal}{ACS Nano} \textbf{\bibinfo{volume}{7}},
	\bibinfo{pages}{9735} (\bibinfo{year}{2013}).
	
	\bibitem[{\citenamefont{Kurtuldu et~al.}(2011)\citenamefont{Kurtuldu, Guasto,
			Johnson, and Gollub}}]{Kurtuldu2011}
	\bibinfo{author}{\bibfnamefont{H.}~\bibnamefont{Kurtuldu}},
	\bibinfo{author}{\bibfnamefont{J.~S.} \bibnamefont{Guasto}},
	\bibinfo{author}{\bibfnamefont{K.~A.} \bibnamefont{Johnson}},
	\bibnamefont{and} \bibinfo{author}{\bibfnamefont{J.~P.}
		\bibnamefont{Gollub}}, \bibinfo{journal}{Proc. Natl. Acad. Sci.} \textbf{\bibinfo{volume}{108}},
	\bibinfo{pages}{10391} (\bibinfo{year}{2011}).
	
	\bibitem[{\citenamefont{Guan et~al.}(2014)\citenamefont{Guan, Wang, and
			Granick}}]{Guan2014}
	\bibinfo{author}{\bibfnamefont{J.}~\bibnamefont{Guan}},
	\bibinfo{author}{\bibfnamefont{B.}~\bibnamefont{Wang}}, \bibnamefont{and}
	\bibinfo{author}{\bibfnamefont{S.}~\bibnamefont{Granick}},
	\bibinfo{journal}{ACS Nano} \textbf{\bibinfo{volume}{8}},
	\bibinfo{pages}{3331} (\bibinfo{year}{2014}).
	
	\bibitem[{\citenamefont{Thorneywork et~al.}(2016)\citenamefont{Thorneywork,
			Aarts, Horbach, and Dullens}}]{Thorneywork2016}
	\bibinfo{author}{\bibfnamefont{A.~L.} \bibnamefont{Thorneywork}},
	\bibinfo{author}{\bibfnamefont{D.~G.} \bibnamefont{Aarts}},
	\bibinfo{author}{\bibfnamefont{J.}~\bibnamefont{Horbach}}, \bibnamefont{and}
	\bibinfo{author}{\bibfnamefont{R.~P.} \bibnamefont{Dullens}},
	\bibinfo{journal}{Soft Matter} \textbf{\bibinfo{volume}{12}},
	\bibinfo{pages}{4129} (\bibinfo{year}{2016}).
	
	\bibitem[{\citenamefont{Chechkin et~al.}(2017)\citenamefont{Chechkin, Seno,
			Metzler, and Sokolov}}]{Chechkin2017}
	\bibinfo{author}{\bibfnamefont{A.~V.} \bibnamefont{Chechkin}},
	\bibinfo{author}{\bibfnamefont{F.}~\bibnamefont{Seno}},
	\bibinfo{author}{\bibfnamefont{R.}~\bibnamefont{Metzler}}, \bibnamefont{and}
	\bibinfo{author}{\bibfnamefont{I.~M.} \bibnamefont{Sokolov}},
	\bibinfo{journal}{Phys. Rev. X} \textbf{\bibinfo{volume}{7}},
	\bibinfo{pages}{21002}
	(\bibinfo{year}{2017}).
	
	\bibitem[{\citenamefont{Chubynsky and Slater}(2014)}]{Chubynsky2014}
	\bibinfo{author}{\bibfnamefont{M.~V.} \bibnamefont{Chubynsky}}
	\bibnamefont{and} \bibinfo{author}{\bibfnamefont{G.~W.}
		\bibnamefont{Slater}}, \bibinfo{journal}{Phys. Rev. Lett.}
	\textbf{\bibinfo{volume}{113}}, \bibinfo{pages}{098302}
	(\bibinfo{year}{2014}).
	
	\bibitem[{\citenamefont{Lanoisel{\'{e}}e and Grebenkov}(2018)}]{Lanoiselee2018}
	\bibinfo{author}{\bibfnamefont{Y.}~\bibnamefont{Lanoisel{\'{e}}e}}
	\bibnamefont{and} \bibinfo{author}{\bibfnamefont{D.~S.}
		\bibnamefont{Grebenkov}}, \bibinfo{journal}{J. Phys. A:
		Math. Theor.} \textbf{\bibinfo{volume}{51}}
	\bibinfo{pages}{145602}
	(\bibinfo{year}{2018}).
	
	\bibitem[{\citenamefont{Malgaretti et~al.}(2016)\citenamefont{Malgaretti,
			Pagonabarraga, and Rubi}}]{Malgaretti2016}
	\bibinfo{author}{\bibfnamefont{P.}~\bibnamefont{Malgaretti}},
	\bibinfo{author}{\bibfnamefont{I.}~\bibnamefont{Pagonabarraga}},
	\bibnamefont{and} \bibinfo{author}{\bibfnamefont{J.}~\bibnamefont{Rubi}},
	\bibinfo{journal}{Entropy} \textbf{\bibinfo{volume}{18}},
	\bibinfo{pages}{394} (\bibinfo{year}{2016}).
	
	\bibitem{Supple}
	See Supplemental Material at [ ] for discussions on sample preparation and calculation of the diffusion constants, emergence of non-Gaussianity with randomness at a higher area fraction of micropillars (28.3\%), trend of variation in the fraction of particles which undergo subdiffusion, and slower and faster particles with randomness \%, two-point hydrodynamic correlation coefficients plot for the samples and match between experimental and predicted $G(\Delta x$) plots. 
	
	\bibitem[{\citenamefont{{D. Allan, T. Caswell, N. Keim, C. van der Wel}}(2018)}]{Allan2018}
	\bibinfo{author}{\bibfnamefont{D.}~\bibnamefont{Allan}},
	\bibinfo{author}{\bibfnamefont{T.}~\bibnamefont{Caswell}},
	\bibinfo{author}{\bibfnamefont{N.}~\bibnamefont{Keim}},
	\bibnamefont{and} \bibinfo{author}{\bibfnamefont{C.}~\bibnamefont{van der Wel}},
	\emph{\bibinfo{title}{{Trackpy
				v0.3.0}}}
	\bibinfo{journal}{Zenodo} \textbf{\bibinfo{volume}{34028}},
	(\bibinfo{year}{2015}).
	
	\bibitem[{\citenamefont{Crocker}(1996)}]{Crocker1996}
	\bibinfo{author}{\bibfnamefont{J. C.}~\bibnamefont{Crocker}},
	\bibinfo{author}{\bibfnamefont{D. G.}~\bibnamefont{Grier}},
	\bibinfo{journal}{J. of Colloid Interface Sci.}
	\textbf{\bibinfo{volume}{179}}, \bibinfo{pages}{298} (\bibinfo{year}{1996}).
	
	\bibitem{He2014}
	K. He, S. T. Retterer, B. R. Srijanto, J. C. Conrad and R. Krishnamoorti, ACS Nano {\bf 8}, 4221 (2014).
	
	
	\bibitem{Hapca2009}
	S. Hapca, J.W. Crawford, and I.M. Young, J. R. Soc. Interface. {\bf 6}, 111 (2009).
	
	\bibitem{Jeon2016}
	J.-H. Jeon, M. Javanainen, H. Martinez-Seara, R. Metzler and I. Vattulainen, Phys. Rev. X  {\bf 6}, 021006 (2016).

	\bibitem{Ghannad2019}
	Z. Ghannad, Phys. Rev. E {\bf 100}, 033211 (2019).
	
	\bibitem{Mora2018}
	S. Mora and Y.  Pomeau, Phys. Rev. E {\bf 98}, 040101(R) (2018).
	
	\bibitem{Jeanneret2016}
	R. Jeanneret, D. O. Pushkin, V. Kantsler and M. Polin, Nat. Comm. {\bf 7}, 12518 (2016).
	
	
	
\end{thebibliography}
%\bibliographystyle{natbib}

%------------------------------------------------

\end{document}

%============================================================
\title{Supplemental Material\par
	Two Coupled Mechanisms Produce Fickian, yet non-Gaussian Diffusion in Heterogeneous Media  } 
%============================================================

\author{Indrani Chakraborty}

\affiliation{School of Chemistry, Tel Aviv
  University, Tel Aviv 6997801, Israel}

\author{Yael Roichman}
\email{roichman@tauex.tau.ac.il}

\affiliation{School of Chemistry, Tel Aviv
  University, Tel Aviv 6997801, Israel}
\affiliation{School of Physics \& Astronomy, Tel Aviv
	University, Tel Aviv 6997801, Israel}

\date{\today}
\pacs{82.70.Dd, %Colloids
	05.40.-a, %Fluctuation phenomena, random processes, noise, and Brownian motion
	47.56.+r}

\maketitle
%------------------------------------------------

\subsection{Preparation of the samples}

The micropillar samples with different degrees of randomness were produced by a standard photolithography technique. At first masks were designed using MATLAB and AUTOCAD. The semi-random masks were obtained via a two step process from the ordered masks. To obtain a mask with a given degree of randomness (say x\%), x \% of the features were chosen randomly and then shifted by random amounts from their positions in the ordered matrix. We used a negative photoresist SU8-3005 to make the micropillars. The micropillars were cylindrical structures 6 $\mu m$ in diameter and 6 $\mu m$ in height. The edge to edge spacing between adjacent pillars were fixed at 6 $\mu m$ for the dilute random case (fractional area coverage or areal density of 20\%) while it was smaller for samples with a higher areal density. Because of the limitations of the photolithography process, at very close proximities, adjacent pillars often touched each other.\par

\subsection{Calculation of diffusion constants}
 Our particles show predominantly Brownian diffusion with diffusion exponent $n$ =1.0, but they still have a small but important fraction with $n$ $\neq$ 1.0. To accurately quantify the diffusion, the diffusion constants were calculated in a different method as compared to taking the slopes from individual $MSD$ vs lag time plots of the particles.
 Following the equation $ \text{{\em MSD}}  = 4\tilde D\tau^n$, $n$ being the diffusion exponent, and $\tilde D$ being the diffusion constant, first straight lines were fitted to the log-log plots of the MSDs, and the intercepts were extracted. Then exponentials of the intercepts were obtained and divided by a factor 4 for 2D diffusion to get the diffusion constants. The diffusion constants therefore have the unit $\mu$m$^2$/s$^n$.

\begin{figure}[h!]
	\centering
	\includegraphics[scale=0.8]{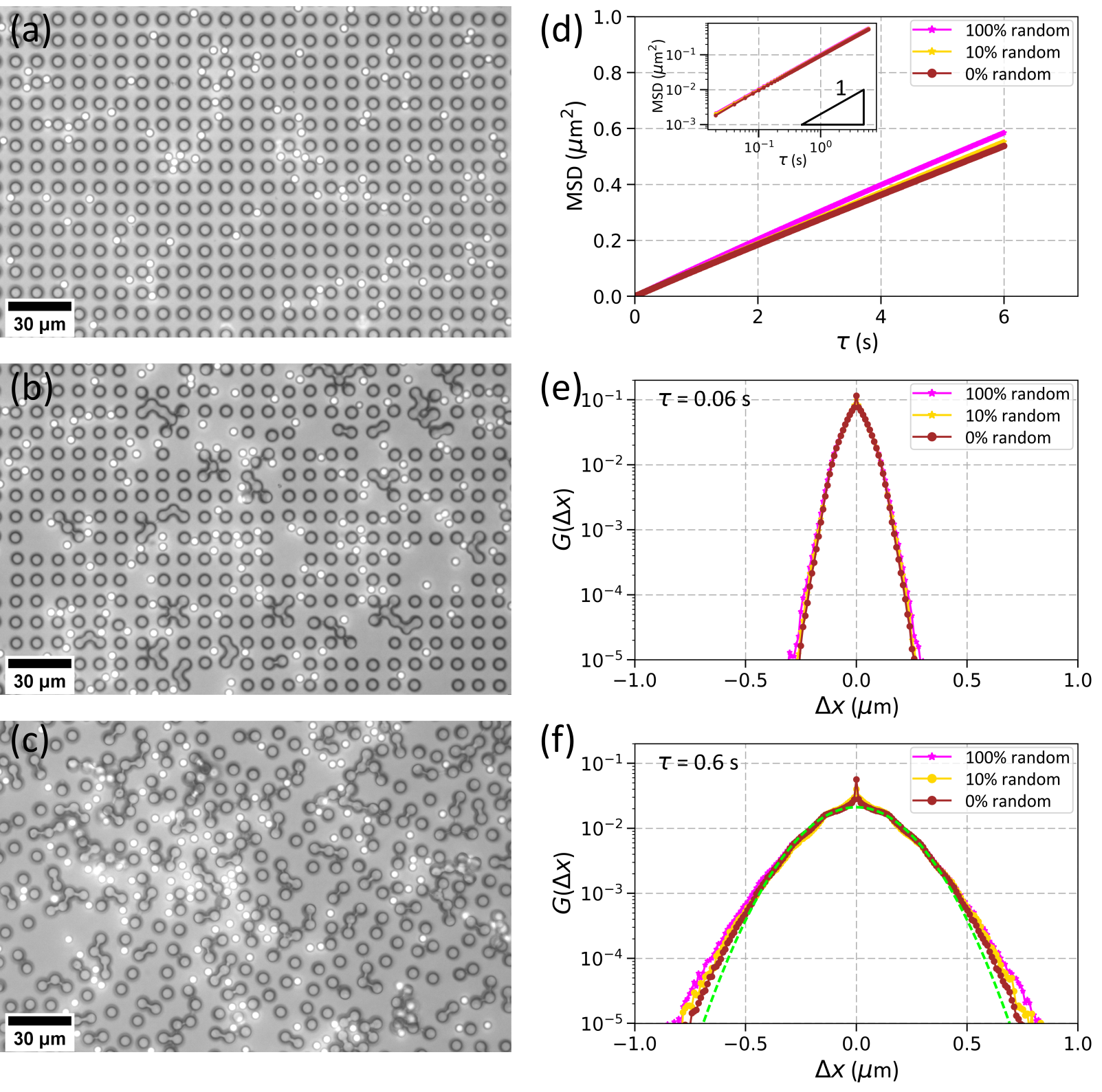}
	\caption{\label{fig:FigS1} Randomness makes diffusion non-Gaussian even at higher area fractions of the micropillars. The microscope images show fluorescent silica particles diffusing in a matrix of micropillars with 28.3\% area fraction in (a) completely ordered (b) semi-random (10\% randomness) and (c) completely random (100\% randomness) arrangements. The MSD plots for randomness values of 0\%, 10\% and 100\% are shown in (d) while inset shows the log-log plot.  $G(\Delta x)$ distribution is shown for lag times (e) $\tau = 0.06s$ and (f) $\tau = 0.6s$. The dashed green line in (f) shows a Gaussian fit to the $G(\Delta x)$ plot for the 100\% random sample. The non-Gaussian parameter $\alpha$ was calculated from the plots. At  $\tau = 1s$, for the 28.3\% area fraction case, $\alpha = 0.20$ for the 0\% random sample and $\alpha = 0.27$ for the 100\% random sample. In contrast for the 20\% area fraction case, the values were $\alpha = 0.13$ for the 0\% random sample and $\alpha = 0.29$ for the 100\% random sample at  $\tau = 1s$. }
\end{figure}

\begin{figure}[h!]
	\centering
	\includegraphics[scale=0.8]{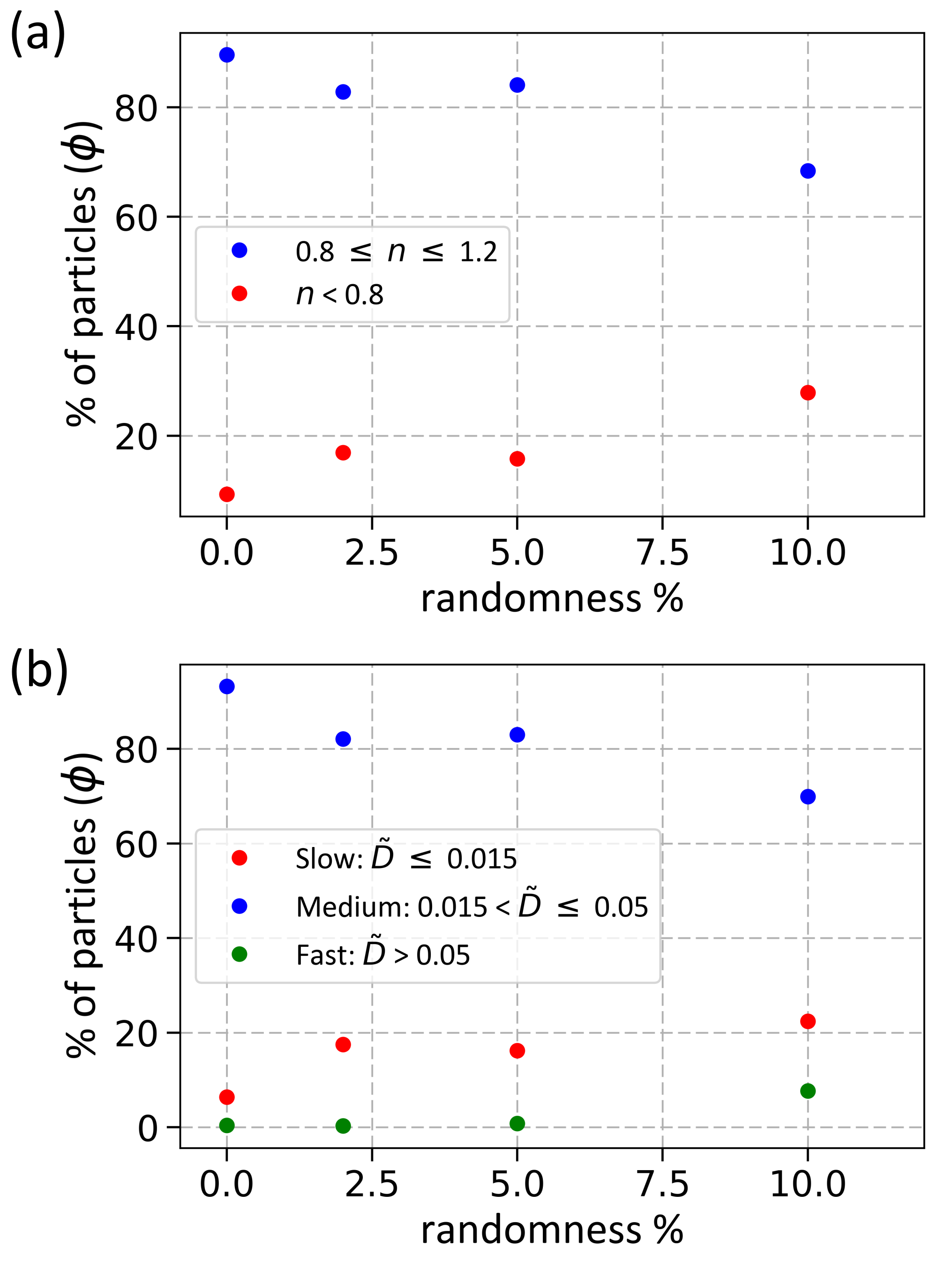}
	\caption{\label{fig:FigS2} (a) How the fraction of individual particles undergoing pure diffusion changes with increasing randomness of the sample. Particles showing pure diffusion are taken to have a diffusion exponent $n$ with 0.8 $\leq $ $n$ $\leq$ 1.2 (shown in blue) whereas particles with n $<$ 0.8 are shown in red.(b) Classifying `fast', `slow' and `medium' particles based on their diffusion constants. Particles with $\tilde D$ $\leq$ 0.015 is referred to as `slow', while particles with 0.015 $<$ $\tilde D$ $\leq$ 0.05 and $\tilde D$ $>$ 0.05 are referred to as `medium' and `fast' particles respectively. Note that the percentage of both fast and slow moving particles increases with increasing structural randomness, while the percentage of medium moving particles decreases. This indicates that the spatial heterogeneities lead to a wider distribution of particle diffusivities with some particles in crowded region or cavities, and some residing in more open areas.}
\end{figure}

\begin{figure}[h!]
	\vspace{2.5cm}
	\centering
	\includegraphics[scale=0.6]{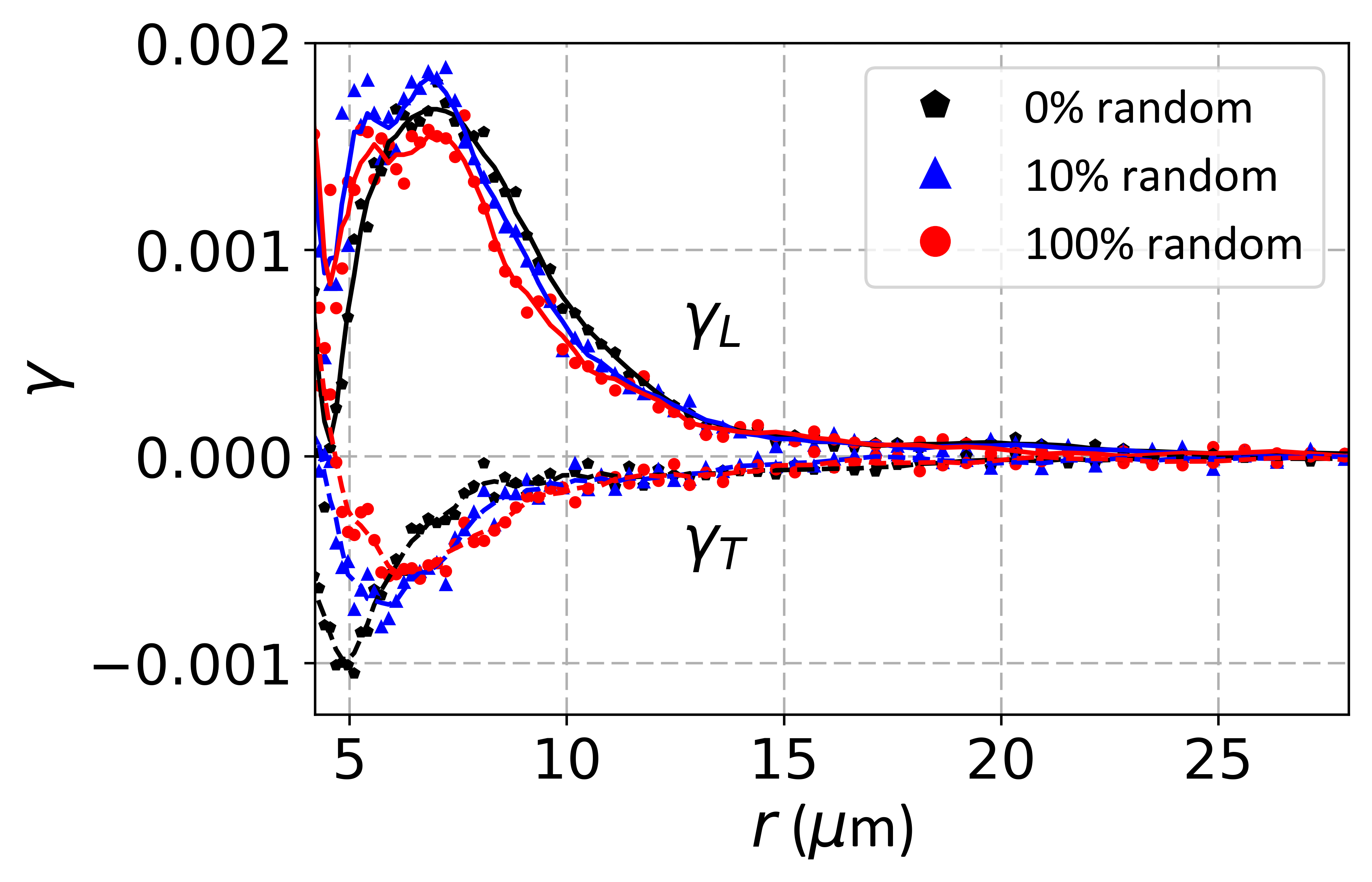}
	\caption{\label{fig:FigS3} Two-point hydrodynamic correlation coefficient ($\gamma$) plot with distance ($r$) for the 0\%, 10\% and 100\% random samples. $\gamma_L$ denote the longitudinal correlation coefficients having positive values (shown by continuous lines) and  $\gamma_T$ denote the transverse correlation coefficients having negative values (shown by dashed lines). They are given by $\gamma_L(r,\Delta{t}) =\langle\Delta{{r_L}^A}(t,\Delta{t}) \Delta{{r_L}^B}(t,\Delta{t})\delta(r-r^{AB})(t)\rangle$ and $\gamma_T(r,\Delta{t}) =\langle\Delta{{r_T}^A}(t,\Delta{t}) \Delta{{r_T}^B}(t,\Delta{t})\delta(r-r^{AB})(t)\rangle$. The lines are drawn by adjacent averaging of datapoints. Note the explicit variation between the correlation coefficients for the three samples at shorter length scales which vanishes at longer length scales.} 
\end{figure}

\begin{figure}[h!]
	\centering
	\includegraphics[scale=0.8]{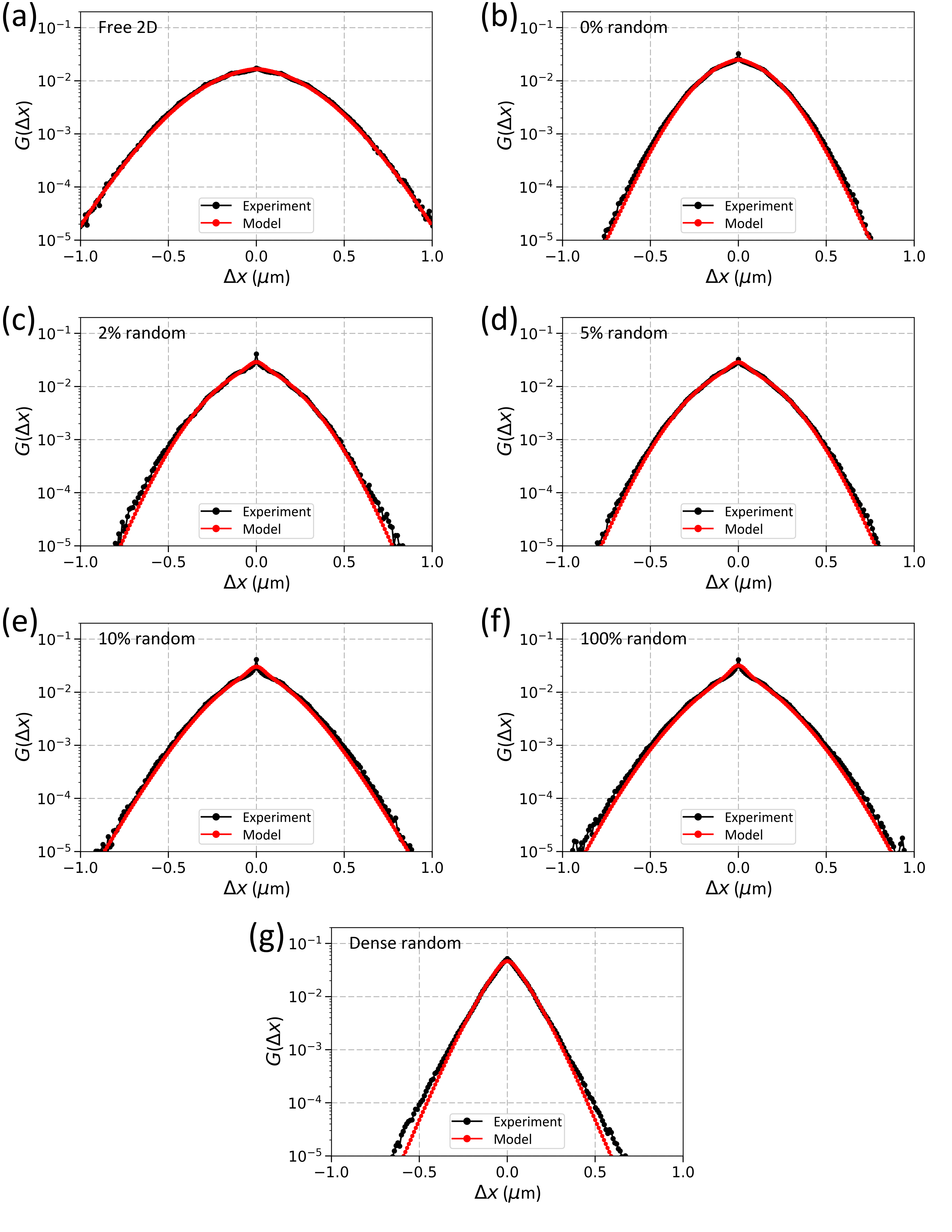}
	\caption{\label{fig:FigS4} Match between experimental and predicted probability distribution of displacements at  $\tau = 0.6 s$. $G(\Delta x$) is obtained from our model by assuming each individual particle having Gaussian displacement distribution and summing over the ensemble with the given weights taken from the measured diffusion constant distributions. $G(\Delta x$) for (a) free 2D, dilute random samples with (b) 0\%, (c) 2\% , (d) 5\%, (e) 10\% and (f) 100\% randomness, and (g) dense random sample is shown. Black denotes the experimental data and red denotes the predicted values from the model. }
\end{figure}